\documentclass[prc]{revtex4}

\usepackage{graphicx}
\usepackage{amsmath}
\begin{document}
\title{Coupling Relativistic Viscous Hydrodynamics to Boltzmann Descriptions}
\author{Scott Pratt}
\affiliation{Department of Physics and Astronomy and National Superconducting Cyclotron Laboratory\\
Michigan State University\\
East Lansing, Michigan 48824}
\author{Giorgio Torrieri}
\affiliation{Institut f\"ur Theoretische Physik\\
J.W. Goethe Universit\"at\\
Max Von Laue Stra\ss e 1, Frankfurt A.M. Germany}
\date{\today}

\begin{abstract}
Models of relativistic heavy ion collisions typically involve both a hydrodynamic module to describe the high density liquid-like phase and a Boltzmann module to simulate the low density break-up phase which is gas-like. Coupling the prescriptions is more complicated for viscous prescriptions if one wants to maintain continuity of the entire stress-energy tensor and currents. Derivations for the viscosity for a gas are reviewed, which then lead to expressions for changes in the phase space occupation based on simple relaxation-time pictures of viscosity. These expressions are shown to consistently reproduce the non-equilibrium components of the stress-energy tensor. An algorithm for generating a Monte Carlo sampling of particles with which to initiate the Boltzmann calculations is also presented.
\end{abstract}


\maketitle

\section{Introduction and Basic Theory}

In the past few years, the community modeling heavy ion collisions has made tremendous progress incorporating viscous effects into hydrodynamic models. However, even viscous models are based on an assumption of short mean free paths and become invalid near the breakup stage when densities become low enough that particles can traverse a significant fraction of the reaction volume. Fortunately, since these densities are low, the system can be treated as a hadronic gas undergoing binary collisions and Boltzmann treatments are justified. This paper focuses on issues related to the interface between the two treatments. For transitions from ideal hydrodynamics, the transition is treated by generating particles consistent with a thermal distribution as a source term for the Boltzmann transport. Here, we present and test a detailed prescription for consistently incorporating viscous effects into such an interface. We also address the question of whether bulk viscosity plays a role in a low density hadronic gas.

From a phenomenological point of view, an understanding of these issues is crucial before the viscosity of the so-called ``perfect fluid'' created in heavy ion collisions can be properly ascertained. Most of the elliptic flow observed in the final state is generated in the earlier stages of the collision \cite{olli} whic can be undestood with ideal hydrodynamics. However, there are substantial differences between ideal and viscous hydrodynamics. Only $50\%$ of the difference of elliptic flow between ideal and viscous hydrodynamics comes from the reduction of the second component of the flow gradient by viscous forces. The rest comes from a modification of the Cooper-Frye \cite{cf} formula at freeze-out \cite{heinz,teavisc}. 
While this modification is generally demanded by energy-momentum conservation in the presence of a pre-existing flow gradient, non-equilibrium effects {\em at freeze-out} are now known \cite{Dusling:2009df} to be capable of modifying the  $v_2$ resulting from a given flow gradient by parametrically non-negligible factors.
Thus, the detailed dynamics of the decoupling of the fluid into particles, and their subsequent interactions, introduce a $50\%$ systematic effect in any determination of $\eta/s$ from comparison of hydrodynamic simulations with data, and needs to be properly accounted for. 
 
A hydrodynamic model provides the fluid velocity $u^\alpha$, the rest-frame energy density $\epsilon$ and charge densities $\vec{\rho}$. Equivalently, $\epsilon$ and $\vec{\rho}$ can be represented by a temperature $T$ and chemical potentials $\vec{\mu}$. Assuming the transition density is sufficiently low to warrant treatment as a thermalized gas, the phase space densities of the various species are straight-forward to generate. Viscous treatments introduce new quantities describing the spatial components of the stress-energy tensor in the fluid frame. Whereas the components are $T_{ij}=P\delta_{ij}$ at equilibrium, where $P(\epsilon,\vec{\rho})$ is the pressure, $T_{ij}$ has additional terms in a viscous treatment. In Navier-Stokes treatments, the additional terms are uniquely determined by the local velocity gradient,
\begin{eqnarray}
T_{ij}&=&P\delta_{ij}+\pi^{(s)}_{ij}+\pi^{(b)}\delta_{ij},\\
\nonumber
\pi^{(s)}_{ij}&=&-\eta\left[\partial_iv_j+\partial_jv_i-(2/3)\delta_{ij}\nabla\cdot v\right],~~~~
\pi^{(b)}=-\zeta\nabla\cdot v,
\end{eqnarray}
where $\eta$ and $\zeta$ are the shear and bulk viscosities respectively. In Israel-Stewart approaches the shear and bulk corrections, $\pi^{(s)}\equiv T_{ij}-({\rm Tr}~T_{ij})\delta_{ij}/3$ and $\pi^{(b)}\equiv({\rm Tr}~T_{ij})/3-P$, are treated as dynamic objects which relax toward Navier-Stokes values \cite{israelstewart,Muronga:2006zw,Muronga:2006zx,Baier:2006um,songheinz}. A consistent interface needs to maintain the continuity of $\pi^{(s)}$ and $\pi^{(b)}$ as well as the densities and collective velocities.

In the next two sections, we present calculations of the bulk and shear viscosities for low density hadronic gases in terms of collision times. Although the shear calculation is a familiar result \cite{prakash,gavin}, the bulk calculation is more complicated, but similar calculations can also be found in the literature \cite{Wiranata:2009cz}. Sophisticated calculations of the shear and bulk viscosity would account for differing relaxation times for different species or for particles with different momentum \cite{mix1}, something which could give rather non-trivial results close to the Hagedorn temperature \cite{mix2}. By using simulations, calculations of the shear viscosity can more accurately incorporate effects from memory being transferred to colliding partners \cite{Demir:2008zz}. The calculations of the next section are not meant to be competitive with these more realistic treatments. Instead, these simple relaxation-time based calculations are included to provide background and context for the algorithms presented here to alter phase-space densities so that they are microscopically consistent with the relaxation time picture. By adjusting the relaxation time, one can reproduce any desired viscosity.

Although the bulk viscosity is zero for a gas of massless particles, or for a gas of non-relativistic particles, a small and nearly negligible bulk viscosity ensues for a gas of semi-relativistic particles. For a gas with a mixture of non-relativistic and highly relativistic particles, e.g. pions and nucleons, one finds a larger, but still small, bulk viscosity. Identical results are derived from the perspective of Kubo relations, and from a more microscopic picture based on the evolving phase space density in the presence of velocity gradients. The next two sections present results for the bulk and shear viscosities respectively, and the subsequent section describes a consistent design of an interface between hydrodynamic and Boltzmann models. The final section summarizes the findings.

\section{The bulk viscosity of a dilute gas}\label{sec:bulk}

The bulk viscosity will be non-zero whenever the trace of the stress-energy tensor, $\sum_i T_{ii}/3$, can differ from the equilibrium pressure, $P(\epsilon,\vec{\rho})$. The inability to maintain equilibrium is assumed to derive from rapidly changing densities, i.e., $\nabla\cdot v\ne 0$. For a non-interacting gas of non-interacting particles,
\begin{eqnarray}
\epsilon&=&\sum_\ell \frac{1}{(2\pi)^3}\int d^3p~E({\bf p},m_\ell) f_{\ell}({\bf p}),\\
\nonumber
\rho_\ell&=&\frac{1}{(2\pi)^3}\int d^3p~ f_{\ell}({\bf p}),\\
\nonumber
T_{ij}&=&\sum_\ell \frac{1}{(2\pi)^3}\int d^3p~\frac{p_ip_j}{E({\bf p},m_\ell)} f_{\ell}({\bf p}),
\end{eqnarray}
where $f_\ell({\bf p})$ is the phase space density for particles of species $\ell$. For massless particles, $E({\bf p},m_\ell)=|p|$, and, as long as the gas is isotropic ($f({\bf p})$ depends on just $|p|$ and not on the angle of {\bf p} with a preferred direction), one finds 
\begin{equation}
\sum_iT_{ii}/3=\frac{\epsilon}{3},
\end{equation}
regardless of the form for $f({\bf p})$. Thus, there is no bulk viscosity for massless gases as the effective pressure is set as a function of the energy density, and is independent of kinetic thermalization. Similarly for non-relativistic gases, $E({\bf p},m_\ell)=m+|p|^2/2m$, and
\begin{equation}
\sum_iT_{ii}/3=\frac{2}{3}(\epsilon-\sum_\ell m_\ell\rho_\ell).
\end{equation}
Again, this depends only on the energy and particle densities, and is independent of whether $f_\ell({\bf p})$ is in an equilibrated form and the gas has zero bulk viscosity. A gas of semi-relativistic particles can have $\sum_iT_{ii}/3\ne P$ by distributing the energy more or less amongst the higher or lower momentum modes. Such a gas with $T\sim m$ can support a small, but negligible for heavy ion collisions, bulk viscosity. The results above are common knowledge in the relativistic heavy ion community. What is less known is that a modest bulk viscosity ensues for a gas with a mixture of non-relativistic and relativistic particles. This will be illustrated below.

For interacting systems, significant bulk viscosities can arise from several non-equilibrium effects \cite{paech,mebulk1}. Near a phase transition, where the vacuum expectations of fields may change suddenly, the finite equilibration time can lead to a peak in the bulk viscosity near $T_c$. Hints of this behavior can be seen in lattice results \cite{kharzeev,mit} and AdS/CFT calculations \cite{ads1,ads2}, and the peak in viscosity could well have a significant effect on subsequent dynamics \cite{mebulk2,krishna,denicol}. Long-range correlations from finite-range interactions or structures, e.g. polymers, experience frictional heating in an isotropic expansion and thus have bulk viscosities. Non-equilibrium chemistry can also lead to a bulk viscosity, but only if one compares $\sum_iT_{ii}/3$ to the pressure calculated with the equilibrium density. If the densities of various species are calculated dynamically, and if non-equilibrium concentrations are then used to calculate the pressure, there is no need to add a bulk viscosity to account for the non-equilibrium pressure. Similarly, if non-equilibrium mean fields are modeled dynamically \cite{paechdumitru} the bulk viscosity associated with non-equilibrium fields can be ignored. In fact, incorporating the effects of non-equilibrium chemistry or fields through a bulk viscosity is potentially clumsy as the effects can easily become non-linear. For this study, we ignore the effects mentioned in this paragraph and consider only the case of a dilute gas where the viscosities are solely due to kinetic non-equilibrium. These effects are much smaller than those mentioned above, and should be the dominant sources at low density when particle interactions are of a purely binary character.

We present calculation of the bulk viscosity of a dilute gas from two perspectives. For each calculation we neglect Bose/Fermi effects and assume that relaxation can be described with a single relaxation time independent of the particle's momentum or species type. This is certainly unrealistic, but the single-relaxation time derivation allows one to precisely relate the alteration to the phase-space density, $\delta f({\bf p})$, due to a change in the stress-energy tensor, $\delta T_{ij}$. First, we calculate the bulk viscosity through the Kubo relation for a dilute gas. This is done by first finding the fluctuation of the pressure for non-interacting particles, then multiplying by the relaxation time. The second calculation is based on a more dynamical picture based on the evolving phase space density in an isotropically expanding medium. Both calculations yield the same result.

\subsection{Calculating the bulk viscosity through the Kubo relation}
The Kubo relation for bulk viscosity,
\begin{eqnarray}
\label{eq:kubobulk}
\zeta&=&(\beta/2)\int d^4x\langle \delta\bar{T}(x)\delta\bar{T}(0)\rangle,\\
\delta \bar{T}&=&\frac{T_{xx}+T_{yy}+T_{zz}}{3}-P,
\end{eqnarray}
assumes the averaging is for states with fixed energy and fixed charges. ``Charges'' refer to anything that is conserved on the time scale for which the correlation lasts. Unfortunately, it is easier to calculate thermal averages in the grand canonical ensemble, which allows energy and charges to fluctuate. For the GC ensemble, one needs to replace \cite{paech}
\begin{equation}
\delta\bar{T}\rightarrow \delta\bar{T}
-\left.\frac{\partial P}{\partial \epsilon}\right|_{\vec{\rho}}\delta T_{00}
-\sum_\ell\left.\frac{\partial P}{\partial\rho_\ell}\right|_{T_{00},\rho_{\ell'\ne \ell}}\delta\rho_\ell.
\end{equation}

The partial derivatives are complicated given that the pressure is most easily calculated in the GC ensemble as a function of the temperature $\beta=1/T$ and the chemical potentials for each species, $\alpha_\ell\equiv-\mu_\ell/T$. After some algebra, they can be expressed in terms of $\partial_\beta P$ and $\partial_{\alpha_\ell}P$. Since we consider only non-interacting hadrons, with no Bose or Fermi statistics, the expression simplifies by using $P_\ell=\rho_\ell T$, and $\partial_\beta P=-(P+\epsilon)/\beta$,
\begin{eqnarray}
\label{eq:dpde}
\left.\frac{\partial P}{\partial \epsilon}\right|_{\rho}&=&
-\frac{PT}{D},\\
\nonumber
\left.\frac{\partial P}{\partial\rho_\ell}\right|_{\epsilon,\rho_{\ell'\ne \ell}}&=&
T+\frac{\epsilon_\ell}{\rho_\ell}\frac{PT}{D},\\
\label{eq:Ddef}
D&\equiv&\sum_\ell(\epsilon_\ell^2/\rho_\ell)+\partial_\beta\epsilon.
\end{eqnarray}
Assuming simple exponential decays in time, characterized by a common relaxation time, the Kubo relation can be calculated once one knows the equal-time fluctuations, $\langle \delta A\delta B\rangle$, where $A$ and $B$ could be any combination of the operators $\delta\bar{T}$, $\delta T_{00}$ or $\delta\rho_\ell$. For non-interacting particles, the operators are of the form,
\begin{eqnarray}
\bar{T}&=&\frac{1}{V}\sum_k \frac{1}{3} \frac{p_k^2}{E_k}\\
\nonumber
T_{00}&=&\frac{1}{V}\sum_k E_k,\\
\nonumber
\rho_i&=&\frac{1}{V}\sum_k,
\end{eqnarray}
where $k$ sums over all particles in the subvolume $V$. For non-interacting particles, equal-time fluctuations of the type $\langle \delta A\delta B\rangle$, which have sums of the type $\sum_{i,j}$ are simplified by eliminating all $i\ne j$ terms, since different particles are uncorrelated and are thus unlikely to be in the same subvolume. The fluctuation of $\bar{T}$ is then,
\begin{equation}
\int d^3x\langle \delta\bar{T}(0)\delta\bar{T}(x_0=0,\vec{x})\rangle
=\frac{1}{V}\sum_k\frac{1}{9} \frac{p_k^4}{E_k^2},
\end{equation}
with similar expressions for the other sums. One can then replace the sums with integrations over phase space with the phase space occupations for the species $\ell$ being $f_\ell({\bf p})=e^{-\beta E_p-\alpha_\ell}$. In terms of the inverse temperature, $\beta$, and the scaled chemical potentials, $\alpha_\ell=-\mu_\ell/T$, the expressions for the densities and correlators are then,
\begin{eqnarray}
\langle\rho_\ell\rangle &=&\int \frac{d^3p}{(2\pi)^3} e^{-\beta E({\bf p},m_\ell)-\alpha_\ell},\\
\nonumber
&=&\frac{e^{-\alpha_\ell}}{(2\pi^2)}\left\{m_\ell^2TK_1(m_\ell/T)+2m_\ell T^2K_2(m_\ell/T)\right\},\\
\langle P\rangle&=&\sum_\ell \int \frac{d^3p}{(2\pi)^3} \frac{p_2}{3E({\bf p},m_\ell)}e^{-\beta E({\bf p},m_\ell)-\alpha_\ell}\\
\nonumber
&=&\sum_\ell \langle\rho_\ell\rangle T,\\
\langle T_{00}\rangle&=&\sum_\ell \int \frac{d^3p}{(2\pi)^3} E({\bf p},m_\ell) e^{-\beta E({\bf p},m_\ell)-\alpha_\ell}\\
\nonumber
&=&\sum_\ell \frac{e^{-\alpha_\ell}}{(2\pi^2)}\left\{m_\ell^3TK_1(m_\ell/T)+3m_\ell^2T^2K_2(m_\ell/T)\right\}\\
\label{eq:dt00dt00}
\int d^3x\langle\delta T_{00}(0)\delta T_{00}(x_0=0,\vec{x})\rangle&=&
\sum_\ell\int \frac{d^3p}{(2\pi)^3} e^{-\beta E({\bf p},m_\ell)-\alpha_\ell} E({\bf p},m_\ell)^2,\\
\nonumber
&=&\sum_\ell\frac{1}{2\pi^2}e^{-\alpha_\ell}\left\{
m_i^4TK_2(m_\ell/T)+m_\ell^3T^2K_3(m_\ell/T)\right\},\\
\label{eq:dtbardtbar}
\int d^3x\langle\delta\bar{T}(0)\delta\bar{T}(x_0=0,\vec{x})\rangle&=&
\sum_\ell\int \frac{d^3p}{(2\pi)^3} e^{-\beta E({\bf p},m_\ell)-\alpha_\ell} \frac{p^4}{9E({\bf p},m_\ell)^2},\\
\nonumber
&&\hspace*{-80pt}=\int d^3x\langle\delta T_{00}(0)\delta T_{00}(x_0=0,\vec{x})\rangle
-\frac{2}{9}\sum_\ell m_\ell^2\langle\rho_\ell\rangle\\
\nonumber
&+&\sum_\ell\frac{m_\ell^4T}{18\pi^2}e^{-\alpha_\ell-m_\ell/T}\left\{1
+\sum_{n=1}^\infty \gamma_n(m_\ell/T)^{2n}\Gamma(-2n+1,m_\ell/T),
\right\},\\
\nonumber&&\gamma_1=-1/2,~~~~\gamma_n=\gamma_{n-1}(n-3/2)/n\\
\int d^3x\langle\delta \rho_\ell(0)\delta \rho_{\ell'}(x_0=0,\vec{x})\rangle&=&
\sum_\ell\int \frac{d^3p}{(2\pi)^3} e^{-\beta E({\bf p},m_\ell)-\alpha_\ell}\delta_{\ell,\ell'}=\langle\rho_\ell\rangle\delta_{\ell,\ell'},\\
\int d^3x\langle\delta \bar{T}(0)\delta T_{00}(x_0=0,\vec{x})\rangle&=&
\sum_\ell\int \frac{d^3p}{(2\pi)^3} e^{-\beta E({\bf p},m_\ell)-\alpha_\ell} \frac{p^2}{3}\\
\nonumber
&=&\int d^3x\langle\delta T_{00}(0)\delta T_{00}(x_0=0,\vec{x})\rangle/3-\sum_\ell m_\ell^2\langle\rho_\ell\rangle/3,\\
\int d^3x\langle\delta \bar{T}(0)\delta \rho_\ell(x_0=0,\vec{x})\rangle&=&
\int \frac{d^3p}{(2\pi)^3} e^{-\beta E({\bf p},m_\ell)-\alpha_\ell} E({\bf p},m_\ell)=\langle T_{00,\ell}\rangle,\\
\int d^3x\langle\delta \bar{T}(0)\delta \rho_\ell(x_0=0,\vec{x})\rangle&=&
\int \frac{d^3p}{(2\pi)^3} e^{-\beta E({\bf p},m_\ell)-\alpha_\ell} E({\bf p},m_\ell)=\langle \bar{T}_\ell\rangle.
\end{eqnarray}
These expressions can then be inserted into an expression for the equal time commutators relevant for the bulk viscosity. After lengthy manipulations using Eq. (\ref{eq:dpde}) along with the expressions for the fluctuations above,
\begin{eqnarray}
\sigma_\zeta^2&\equiv&\frac{1}{V}\langle \left[\int d^3x \left(\delta\bar{T}
-\left.\frac{\partial P}{\partial \epsilon}\right|_{\vec{\rho}}\delta T_{00}
-\sum_\ell\left.\frac{\partial P}{\partial\rho_\ell}\right|_{T_{00},\rho_{\ell'\ne \ell}}\delta\rho_\ell\right)\right]^2
\rangle\\
&=&\int d^3x \langle\delta\bar{T}(0)\delta\bar{T}(x)\rangle-P\left(1+\left.\frac{\partial P}{\partial\epsilon}\right|_{\vec{\rho}}\right),\\
\left.\frac{\partial P}{\partial\epsilon}\right|_{\vec{\rho}}&=&\frac{-PT}{\sum_\ell(\epsilon_\ell^2/\rho_\ell)+\partial_\beta\epsilon} \label{sigmadef}
\end{eqnarray}
where $\partial_\beta\epsilon=-\int d^3x\langle\delta T_{00}(0)\delta T_{00}(x)\rangle$ is given by Eq. (\ref{eq:dt00dt00}) and is related to the specific heat, $\partial_\beta\epsilon=-T^2\partial_T\epsilon$.
If the relaxation time $\Delta\tau$ is independent of momentum and relaxation time, the Kubo relation for the bulk viscosity, Eq. (\ref{eq:kubobulk}), becomes
\begin{eqnarray}
\label{eq:zetaresult}
\zeta&=&\beta\sigma_\zeta^2\Delta\tau=\left[\beta \int d^3x \langle\delta\bar{T}(0)\delta\bar{T}(x_0=0,\vec{x})\rangle
-P\left(1+\left.\frac{\partial P}{\partial\epsilon}\right|_{\vec{\rho}}\right)
\right]\Delta\tau,\\
\nonumber
\left.\frac{\partial P}{\partial\epsilon}\right|_{\vec{\rho}}&=&\frac{-PT}{\sum_\ell(\epsilon_\ell^2/\rho_\ell)+\partial_\beta\epsilon},
\end{eqnarray}
with the expectation $\int d^3x \langle\delta\bar{T}(0)\delta\bar{T}(x)\rangle$ and $\partial_\beta\epsilon=-\int d^3x\langle\delta T_{00}(0)\delta T_{00}(x)\rangle$ being given in Eq.s (\ref{eq:dt00dt00}-\ref{eq:dtbardtbar}). 

\subsection{Alternate derivation of $\zeta$}

The bulk viscosity can also be understood by calculating $\bar{T}$ for a gas expanding under an isotropic velocity gradient described by $\nabla\cdot v$ for a relaxation time $\Delta\tau$ without collisions, and compare it to the pressure one would find if equilibrium had been maintained. For the collision-less expansion, one can consider a Hubble expansion where the collective velocities are $v_i=x_i/t$. For a particle whose momentum was ${\bf p}_0$ at proper time $\tau_0=\sqrt{t_0^2-x_0^2}$, the momentum at proper time $\tau$ as measured in the local rest frame, will be, in the absence of collisions,
\begin{equation}
{\bf p}={\bf p}_0\frac{\tau_0}{\tau},
\end{equation}
and for small time differences,
\begin{equation}
\Delta{\bf p}=-{\bf p}\frac{\Delta\tau}{\tau},
\end{equation}
and for the Hubble expansion, $\nabla\cdot v=3/\tau$, so that
\begin{equation}
\Delta{\bf p}=-{\bf p}(\nabla\cdot v)\Delta\tau/3.
\end{equation}
One can then express the pressure and change in pressure as:
\begin{eqnarray}
\bar{T}&=&\frac{1}{V}\sum_\ell\int \frac{Vd^3p}{(2\pi)^3}e^{-\beta E({\bf p},m_\ell)-\alpha_\ell}\frac{p^2}{3E_p},\\
\nonumber
\Delta\bar{T}&=&-\frac{\Delta V}{V}\bar{T}+\sum_\ell\int \frac{d^3p}{(2\pi)^3} e^{-\beta E({\bf p},m_\ell)-\alpha_\ell}\Delta\frac{p^2}{9E({\bf p},m_\ell)}.
\end{eqnarray}
This last step exploited the Liousville theorem stating that $dN=Vd^3p f(p)/(2\pi)^3$, where $f$ is the phase space density, stays constant. One can then use the above expression for $\Delta{\bf p}$ along with the fact that 
$\Delta V=V(\nabla\cdot v)\Delta\tau$, to obtain,
\begin{eqnarray}
\Delta\bar{T}_{\rm no~coll.}&=&-(\nabla\cdot v)\Delta\tau
\left[P+\sum_\ell\int \frac{d^3p}{(2\pi)^3} e^{-\beta E({\bf p},m_\ell)-\alpha_\ell}
{\bf p}\cdot\nabla_p \frac{p^2}{9E({\bf p},m_\ell)}\right],\\
\nonumber
&=&-(\nabla\cdot v)\Delta\tau\left[P+\sum_\ell\int \frac{d^3p}{(2\pi)^3} e^{-\beta E({\bf p},m_\ell)-\alpha_\ell}
\left(\frac{2p^2}{9E({\bf p},m_\ell)}-\frac{p^4}{9E({\bf p},m_\ell)^3}\right)\right].
\end{eqnarray}
The last term in the integral can be re-expressed as an integral over energy, $pdp\rightarrow E_pdE_p$, and after integrating by parts one finds
\begin{eqnarray}
\Delta\bar{T}_{\rm no~coll.}&=&-(\nabla\cdot v)\Delta\tau\left[\beta\sum_\ell\int\frac{d^3p}{(2\pi)^3}e^{-\beta E({\bf p},m_\ell)-\alpha_\ell}\frac{p^4}{9E({\bf p},m_\ell)^2}\right]\\
\nonumber
&=&-(\nabla\cdot v)\Delta\tau\int d^3x~\left\langle \delta\bar{T}(0)\delta\bar{T}(\vec{x})\right\rangle,
\end{eqnarray}
with the integral being given in Eq. (\ref{eq:dtbardtbar}). To calculate the change if thermal equilibrium were maintained, one can apply conservation of entropy and particle number to find expressions for the changes in energy density and particle density during a time $\Delta\tau$,
\begin{eqnarray}
T\Delta S&=&0=\Delta E-P\Delta V,\\
\nonumber
(P+\epsilon)\nabla\cdot v&=&-\partial_t\epsilon\\
\nonumber
\Delta\epsilon&=&-(\nabla\cdot v)\Delta\tau(P+\epsilon),\\
\nonumber
\Delta\rho_\ell&=&-(\nabla\cdot v)\Delta t\rho_\ell.
\end{eqnarray}
The change of the equilibrium pressure during this time is
\begin{eqnarray}
\Delta P_{\rm equil}&=&\left.\frac{\partial P}{\partial\epsilon}\right|_\rho\Delta\epsilon
+\left.\sum_\ell\frac{\partial P}{\partial\rho_\ell}\right|_{\epsilon,\rho_{j\ne \ell}}\Delta\rho_\ell,\\
\nonumber
\Delta P_{\rm equil}&=&-(\nabla\cdot v)\Delta\tau\left\{
\left.\frac{\partial P}{\partial\epsilon}\right|_{\rho}(P+\epsilon)
+\left.\sum_\ell\frac{\partial P}{\partial\rho_\ell}\right|_{\epsilon,\rho_{j\ne \ell}}\rho_\ell\right\}\\
\nonumber
&=&-(\nabla\cdot v)\Delta\tau\left(P+P\left.\frac{\partial P}{\partial\epsilon}\right|_{\rho}\right)
\end{eqnarray}
Taking the difference, the effect of collisions is:
\begin{equation}
\Delta\bar{T}_{\rm no~coll.}-\Delta P_{\rm equil.}
=-(\nabla\cdot v)\Delta\tau\left(-P-P\left.\frac{\partial P}{\partial\epsilon}\right|_\rho
+\beta\int d^3x \langle\bar{T}(0)\bar{T}(x)\rangle\right).
\end{equation}
After applying the definition of bulk viscosity, $\Delta\bar{T}=-\zeta\nabla\cdot v$, and using the expression for $\partial P/\partial\epsilon$ in Eq. (\ref{eq:dpde}), this gives exactly the same expression for $\zeta$ as in Eq. (\ref{eq:zetaresult}) of the previous subsection.
\begin{eqnarray}
\zeta&=&\Delta\tau\left(-P+\frac{P^2T}{D}
+\beta\int d^3x \langle\delta\bar{T}(0)\delta\bar{T}(x)\rangle\right),\\
\nonumber
D&\equiv&\partial_\beta\epsilon+\sum_\ell\frac{\epsilon_\ell^2}{\rho_\ell}.
\end{eqnarray}

\subsection{Temperature dependence of the viscosity}
According to the Kubo relations, the viscosity is a product of an equal-time fluctuation and the relaxation time. The relaxation time, $\Delta \tau$, falls roughly inversely with the density thus making both the shear and bulk viscosities strongly temperature dependent. The equal-time fluctuations also depend on temperature. For the bulk viscosity, the fluctuation, $\int d^3x \langle \delta\bar{T}(0)\delta\bar{T}(x)\rangle$, is zero for both the non-relativistic and ultra-relativistic limits. This can be understood by considering the behavior of the momenta as measured in the local rest-frame,
\begin{equation}
\Delta p_i=-\Delta \tau p_j\frac{\partial v_i}{\partial x_j}.
\end{equation}
For an isotropic velocity gradient,
\begin{equation}
\Delta p_i=-\nabla\cdot v\frac{\Delta\tau}{3}p_i.
\end{equation}
The corresponding change in the phase space density is proportional to 
\begin{equation}
\Delta f\propto \Delta p_i\frac{\partial}{\partial p_i}f=-\beta\frac{p_i^2}{E({\bf p})}f.
\end{equation}
For a change in the temperature and chemical potential,
\begin{equation}
\Delta f\propto -\Delta\beta E_pf-\Delta\alpha f,
\end{equation}
which means that for an ultra-relativistic particle, $E_p=p$, the change in the phase space density is equivalent to a change in the temperature. The phase space occupancy is thus changed in a way consistent with changing the temperature, which means that the distribution remains thermal. Since a thermal distribution is unchanged by collisions, the collisions play no role and the collisionless limit is indistinguishable from the equilibrated limit, thus there is no bulk viscosity. For a flat-space Hubble expansion one finds that the collisionless limit is equivalent to having the temperature fall as $1/\tau$. 

For the non-relativistic limit, $E_p=m+p^2/2m$, and the phase space density falls in such a way that can be perfectly well described by changing the temperature and chemical potential. Again, this is consistent with local kinetic thermalization which makes the collision-less and equilibrated limits indistinguishable, and the bulk viscosity is zero. For the Hubble example, the temperature falls as $1/\tau^2$

However, the equivalence is broken for the case where the temperature is of the order of the mass. The fluctuation is then small, which leads to a small viscosity. A larger fluctuation occurs for the case of a mixture of relativistic and non-relativistic systems. This is most easily understood from the Hubble example above. If one component of the phase space density cools as $1/\tau^2$ while the other cools as $1/\tau$, the two components become characterized by different temperatures \cite{Pratt:1998gt}. This departure from equilibrium enables entropy production. Such is the case for a hadronic gas with temperatures $\gtrsim 100$ MeV, a range for which the pion is relativistic while the baryons are mostly non-relativistic.

\begin{figure}
\centerline{\includegraphics[width=0.45\textwidth]{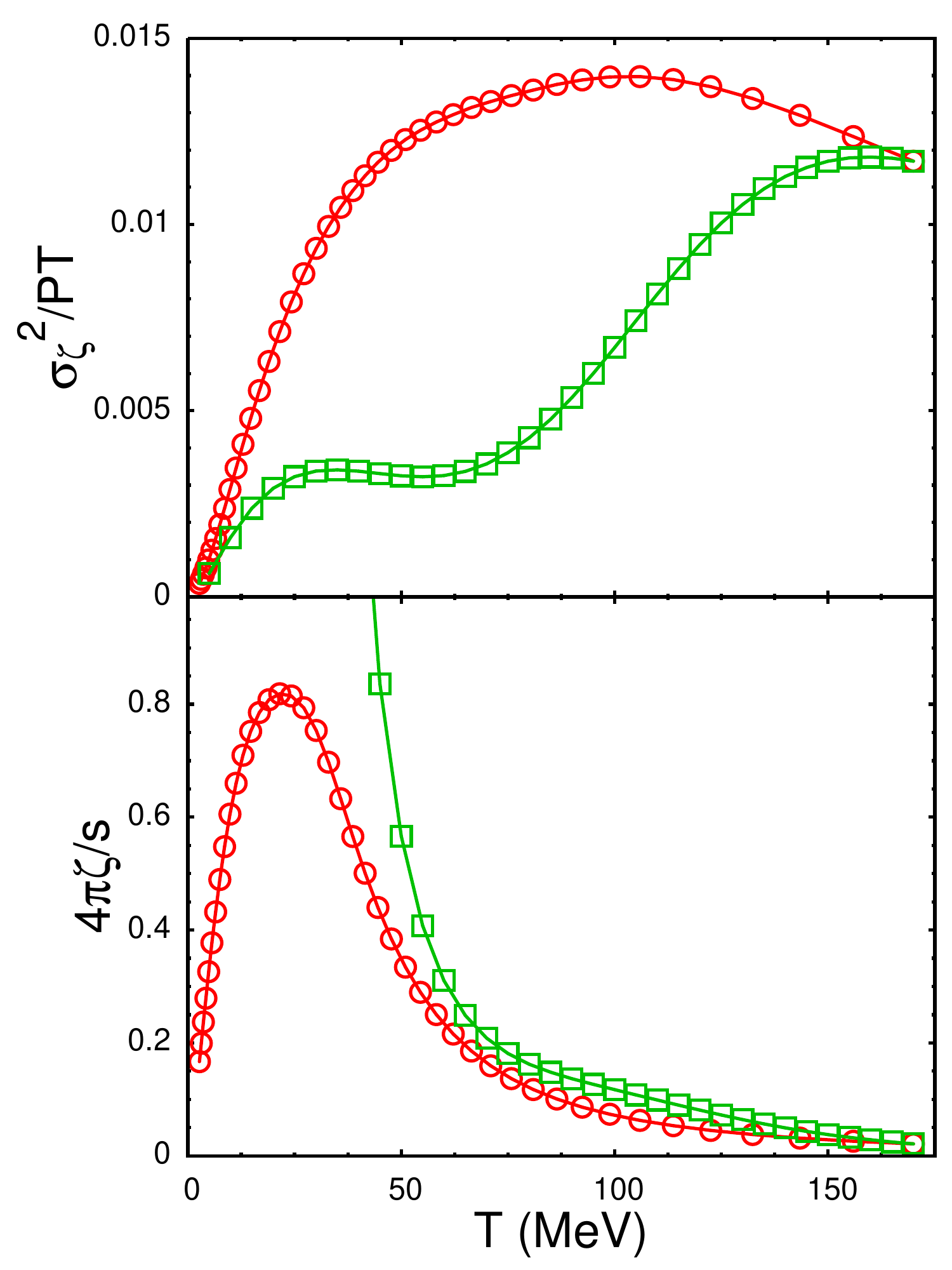}\hspace*{0.02\textwidth}\includegraphics[width=0.45\textwidth]{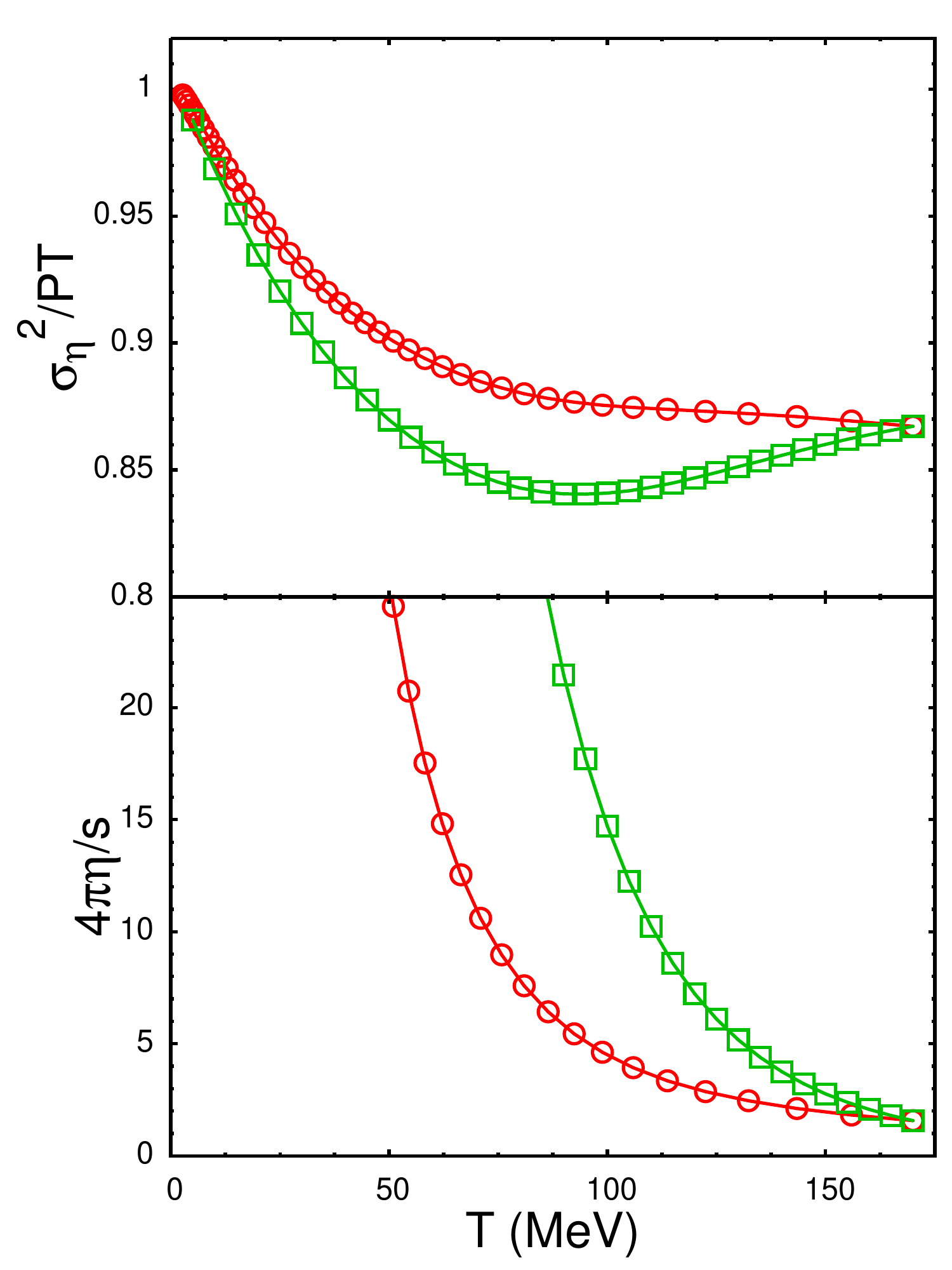}}
\caption{\label{fig:bulk}
Bulk (lower left panel) and shear (lower right panel) viscosities are shown as a function of temperature for two cases of a hadronic gases: a chemically equilibrated gas (squares) and a second gas whose composition is that of a system that isentropically expanded from a temperature of 170 MeV while conserving several effective charges described in the text. The upper panels show the corresponding equal-time fluctuations of the stress-energy tensor elements defined in Eq.s (\ref{eq:kubobulk}) and (\ref{eq:kuboshear}). One reproduces the viscosities by multiplying the relaxation time and the fluctuations, then dividing by $T$. The low values of the bulk viscosity derives from the small fluctuations, which disappear in both the ultra-relativistic and non-relativistic limits, hence the return to zero at low temperature. For temperatures of interest for interfacing hydrodynamic and microscopic treatments, $T\sim 150$ MeV, the bulk viscosities due to kinetic non-equilibrium are negligible. 
}
\end{figure}

Figure \ref{fig:bulk} displays both the fluctuation $\sigma_\zeta^2$ defined as in Eq. \ref{sigmadef} and the bulk viscosity, $\zeta=\beta\sigma_\zeta^2\Delta\tau$, for a hadron gas assuming a that the relaxation time is given by the simple expression, $\Delta\tau=1/\sigma\rho_{\rm tot}$, where $\rho_{\rm tot}$ is the total density of hadrons and $\sigma$ is a fixed cross section of 20 mb. For this energy range cross sections are usually rather larger, but many of the collisions carry momentum forward which should lengthen the effective relaxation time. More detailed microscopic considerations are taken into account in \cite{Wiranata:2009cz}. The hadronic gas is calculated for two cases. First, an equilibrated gas is considered for temperatures below 170 MeV, using only the hadrons from the lowest lying baryon and mesons flavor decuplets and octets. Secondly, a gas is considered under the constraints that the net number of strange particles, baryons, effective pion number, omegas and etas are fixed from what the gas would have had at a temperature of 170 MeV before undergoing an isentropic expansion. In both cases, the fluctuations, scaled by $PT$, are small throughout the temperature range. At low temperatures the fluctuations return to zero as even the pions become non-relativistic. The two cases have very different densities at low temperatures. Due to the chemical non-equilibrium, the non-equilibrated state has many more particles due to the finite fugacities that develop \cite{Greiner:1993jn}, which gives much shorter relaxation times and thus much smaller viscosities. Since the temperature range for which one would switch from hydrodynamics to microscopic prescriptions is in the neighborhood of 150 MeV, the behaviors shown in Fig. \ref{fig:bulk} for lower temperatures are only for academic interest. For either case, the bulk viscosity is negligible in the range of interest, although it should be emphasized that these figures illustrate only the viscosities due to kinetic non-equilibrium.

\section{Calculations of Shear Viscosity}
Expressions for the shear viscosity are more common in the literature and can be found in text books \cite{weinberg}. For completeness, derivations of the shear viscosity, $\eta$, are provided here from the same two perspectives as the previous section. First, the shear viscosity is derived from the Kubo relation, and secondly, it is derived from a more kinematic perspective by considering the alteration of the phase space density in the presence of a velocity gradient. As in the previous section, the two expressions are identical. The latter derivation illustrates how one can choose a phase space density for a microscopic simulation such that it is consistent with a viscous hydrodynamic situation.

Unlike the bulk case, the Kubo relation for the shear viscosity,
\begin{equation}
\label{eq:kuboshear}
\eta=\frac{\sigma_\eta^2\Delta\tau}{T},~~~
\sigma_\eta^2\equiv\int d^3r\left\langle \delta T_{xy}(0)\delta T_{xy}({\bf r})\right\rangle,
\end{equation}
does not have the complications associated with going from the canonical to the microcanonical ensemble, because $T_{xy}$ is not correlated with fluctuations of the energy density or conserved charges. Thus, Eq. (\ref{eq:kuboshear}) can be evaluated in the grand canonical ensemble. As in the previous section, the fluctuation in the dilute limit only comes from a particle being correlated with itself for a time $\Delta\tau$, which combined with the relaxation time leads to
\begin{eqnarray}
\label{eq:shearp2}
\eta&=&\beta\Delta\tau\sum_\ell\int\frac{d^3p}{(2\pi\hbar)^3}\frac{p_x^2p_y^2}{E({\bf p},m_\ell)^2}e^{-\beta E({\bf p},m_\ell)-\alpha_\ell}\\
\nonumber
&=&\frac{\beta}{15}\Delta\tau\sum_\ell\int\frac{d^3p}{(2\pi\hbar)^3}\frac{p^4}{E({\bf p},m_\ell)^2}e^{-\beta E({\bf p},m_\ell)-\alpha_\ell}\\
\nonumber
&=&\frac{3\beta\Delta\tau}{5}\int d^3x\left\langle \delta\bar{T}(0)\delta\bar{T}(x_0=0,\vec{x})\right\rangle,
\end{eqnarray}
with the integral being given in Eq. (\ref{eq:dtbardtbar}).

One can also derive the expression for the shear in Eq. (\ref{eq:shearp2}) from the perspective of how the phase space density changes in the presence of a velocity gradient. To see this one considers a particle of momentum ${\bf p}$, that moves a distance $({\bf p}/E_p)\Delta t$ between collisions. The particle then finds itself in a region whose collective velocity has changed by an amount,
\begin{equation}
\Delta v_{{\rm coll},i}=\frac{\partial v_{{\rm coll},i}}{\partial x_j}
(p_j/E_p)\Delta t.
\end{equation}
For small $\Delta v_{{\rm coll},i}$, the momentum as measured in the rest frame of the new neighbors, is lessened by an amount,
\begin{equation}
\label{eq:Delp}
\Delta p_i=-E_p\Delta v_{{\rm coll},i}=-\frac{\partial v_{{\rm coll},i}}{\partial x_j}
p_j\Delta t.
\end{equation}

In between collisions, the phase space density is fixed for a point in phase space following a particle's trajectory, i.e., Liouville's theorem. This allows one to calculate the stress-energy tensor element,
\begin{eqnarray}
T_{xy}&=&\sum_\ell \int \frac{d^3p}{(2\pi\hbar)^3} f({\bf p}) \frac{p_xp_y}{E({\bf p},m_\ell)}\\
\nonumber
&=&\sum_\ell \int \frac{d^3p'}{(2\pi\hbar)^3} f({\bf p}') \frac{p_xp_y}{E({\bf p},m_\ell)},
\end{eqnarray}
where the Jacobian $d^3p/d^3p'$ from the transformation in Eq. (\ref{eq:Delp}), ${\bf p}={\bf p}'+\Delta{\bf p}$, is unity. Assuming that the only non-zero component of the velocity gradient is $\partial_xv_y$, 
\begin{eqnarray}
\Delta T_{xy}&=&
\sum_\ell \int \frac{d^3p}{(2\pi\hbar)^3} f({\bf p}) \left(-p_y \Delta t\frac{\partial v_x}{\partial y}\right) \frac{\partial}{\partial p_x}\frac{p_xp_y}{E({\bf p},m_\ell)}\\
&=&-\left(\frac{\partial v_x}{\partial y}\right)\beta \Delta t\sum_\ell\int\frac{d^3p}{(2\pi\hbar)^3}
e^{-\beta E({\bf p},m_\ell)-\alpha_\ell} \frac{p_x^2p_y^2}{E({\bf p},m_\ell)^2},
\end{eqnarray}
where the last step involved an integration by parts for the integral over $p_x$. Using the definition of shear viscosity, one can read off an expression for $\eta$,
\begin{equation}
\label{eq:shear}
\eta=\beta\Delta t\sum_\ell\int\frac{d^3p}{(2\pi\hbar)^3}e^{-\beta E({\bf p},m_\ell)-\alpha_\ell}\frac{p_x^2p_y^2}{E({\bf p},m_\ell)^2},
\end{equation}
which is the same expression derived from the Kubo relation, Eq. (\ref{eq:shearp2}).

\section{Interfacing hydrodynamic and Boltzmann modules}
\label{sec:interface}

It is inappropriate to model the later stages of a reaction with hydrodynamics. Once the mean free paths approach a similar scale to the system size, local kinetic equilibrium can be lost between different species. For instance, due to their lighter mass pions begin to flow outward faster than protons \cite{sorge_protonlag}, and due to the fact that they are relativistic, they also cool more slowly \cite{Pratt:1998gt}. Although small deviations can be incorporated into hydrodynamics with viscosities, heat conductivities or particle number diffusions, at low temperatures, it becomes necessary to model with a Boltzmann prescription. The interface temperature needs to be high enough such that viscous hydrodynamics is warranted just above the interface temperature, while it needs to be low enough that hadrons are well defined objects and that the dynamics are not greatly affect by non-binary interactions. Albeit narrow, there does seem to exist such a window with $140\lesssim T\lesssim 165$ MeV.

In order to initialize the Boltzmann description, one needs to choose a phase space density, $f({\bf p},x)$. 
Since $f({\bf p},x)$ depends on the three-dimensional momentum ${\bf p}$ at any space-time point $x$, it can be altered in innumerable ways to incorporate viscous corrections, so long as one maintains a consistency between the stress-energy tensor elements of the input hydrodynamic model and that of the microscopic model,
\begin{equation}\label{eq:consistency}
T_{ij}=\sum_{{\rm species}~\ell}\int\frac{d^3p}{(2\pi)^3}\frac{p_ip_j}{E({\bf p})}
f_{\rm \ell}({\bf p}),
\end{equation}
where $\ell$ refers to particular species (spin and flavor degeneracy factors are suppressed). The functional form for the modification of $f$ might or might not be based on a physical picture. A common ansatz suggested by Grad \cite{grad} (motivated by the expansion of the phase space density in a viscous fluid) for correcting for shear is \cite{teavisc,Dusling:2007gi,Monnai:2009dh,Baier:2006um}
\begin{equation}\label{eq:gradform}
f({\bf p},x)= f_{\rm eq}({\bf p},x)\left[1+C(p)p_ip_j\pi^{(s)}_{ij}\right],
\end{equation}
where $f_{\rm eq}$is the equilibrated phase space density determined by the energy and charge densities. As discussed in \cite{Dusling:2009df}, the momentum dependence of $C(p)$ can be picked to reflect the energy dependence for which particles lose memory regarding their original momentum. Additionally, it might depend on species type, e.g., since protons have larger cross sections than pions, they have relatively smaller viscous corrections to their phase space density.

The advantage of Eq. \ref{eq:gradform} is that it is related in a straight-forward manner to the viscous part of the stress-energy tensor (whether calculated in Navier-Stokes or the more general, dynamical, Israel-Stewart), as it is essentially an estimate of the Grad correction term. A disadvantage of this form in Eq. (\ref{eq:gradform}) is that for high momentum and for certain directions of ${\bf p}$ one finds negative phase space densities.  The unphysical negative phase space density is not surprising: The typical departure of thermalization varies with the momentum of the particle (typically, for $p \gg T$ $\sigma \sim 1/p^2$ so $\Delta \tau \sim p^2$).  Thus, for large momenta, the first term of the gradient expansion will be insufficient.  

For the purpose of this work, we shall consider an alternative prescription, which, as we will see, is less affected by the high-momentum pathological behaviour when implemented in a Monte-Carlo code.   In this prescription,
\begin{equation}
f({\bf p})=\exp\left\{-E({\bf p}')/T'-\alpha'\right\}.
\end{equation}
The primes on the chemical potential and temperature note that the temperatures and chemical potentials can be different than the quantities one would choose to match the energy and charge densities in a non-viscous theory. The momenta $p$ and $p'$ will be related to one another through the relation,
\begin{equation}
\label{eq:linearptrans}
p_i=p'_i+\lambda_{ij}p'_j.
\end{equation}
Note that for $\lambda_{i j} p'_{j} \ll 1$, ie for small momenta, the 
first Taylor coefficient of this formula should lead to Eq. 
\ref{eq:gradform}.

Only for bulk effects, i.e., $\sum_i (T_{ii}-P)\ne 0$, will the temperature $T'$ and chemical potential $\alpha'$ differ from the equilibrated values. An advantage of this form is that it can be applied to Monte Carlo generation of momenta in a straight-forward manner by generating ${\bf p}'$ according to a thermal distribution, then performing the linear transformation to generate the momenta ${\bf p}$.

To motivate the linear behavior in Eq. (\ref{eq:linearptrans}), one can consider a particle with momentum $p'$ at time $\tau_0$. After a small time $\Delta\tau$ which can be thought of as the collision time or the relaxation time, the particle will have moved a distance $\Delta{\bf r}=({\bf p}'/E({\bf p}'))\Delta\tau$. At that time, it will have moved to a region with velocity $\Delta v_i=(\partial v_i/\partial r_j)\Delta r_j$. The momentum as measured in the new frame will be:
\begin{equation}
p_i=p'_i-E({\bf p}')\Delta v_i=p'_i-E({\bf p}')\frac{\partial v_i}{\partial r_j}\frac{p_j}{E({\bf p}')}\Delta\tau
=p'_i-\Delta\tau\partial_jv_i p'_j,
\end{equation}
and
\begin{equation}
\label{eq:lambdaguess}
\lambda_{ij}=-\Delta\tau\partial_jv_i.
\end{equation}
For irrotational flow, the Navier Stokes equation assumes that the velocity gradient is proportional to the viscous correction to the stress-energy tensor. In particular, we will assume
\begin{equation}
\label{eq:Adef}
\lambda_{ij}=A^{(s)}\pi^{(s)}_{ij}+A^{(b)}\pi^{(b)}\delta_{ij},
\end{equation}
and will find the coefficients $A^{(s)}$ and $A^{(b)}$ that satisfy the consistency requirements of Eq. (\ref{eq:consistency}) in the limit that $\lambda$, or equivalently $\pi^{(s)}$ and $\pi^{(b)}$, are small.

With this transformation, one can apply the Liouville theorem, $f({\bf p})=f({\bf p}')$,
\begin{equation}
T_{ij}=\sum_{\ell}e^{\alpha'_\ell}\int \frac{d^3p'}{(2\pi)^3} e^{-E({\bf p}')/T'}\frac{p_ip_j}{E({\bf p})},
\end{equation}
where it has been assumed that the volume is fixed. The stress-energy tensor changes due to $\lambda$ and due to $\Delta T=T'-T$ and $\Delta\alpha=\alpha'-\alpha$,
\begin{eqnarray}
\label{eq:Tijprime}
\pi^{(s)}_{ij}&=&\sum_{\ell}e^{-\alpha_\ell}\int\frac{d^3p'}{(2\pi)^3}e^{-E({\bf p}')/T}\left\{
\frac{2p'_i\lambda^{(s)}_{jk}p'_k}{E({\bf p}')}-\frac{p'_ip'_j}{E({\bf p}')^3}p'_k\lambda^{(s)}_{km}p'_m
\right\},\\
\nonumber
\pi^{(b)}&=&\sum_{\ell}e^{-\alpha_\ell}\int\frac{d^3p'}{(2\pi)^3}e^{-E({\bf p}')/T}\left\{
\frac{2\lambda^{(b)}p^{\prime 2}}{3E({\bf p}')}-\frac{p^{\prime 4}}{3E({\bf p}')^3}\lambda^{(b)}
\right\}\\
\nonumber
&+&\sum_\ell e^{-\alpha_\ell}\int\frac{d^3p'}{(2\pi)^3}e^{-E({\bf p}')/T}\frac{p'^2}{3E({\bf p}')}\left\{
-\Delta\alpha_\ell+\frac{E({\bf p}')\Delta T}{T^2}\right\}.
\end{eqnarray}
Here, $\lambda^{(s)}$ is the traceless part of $\lambda$ and $\lambda^{(b)}$ describes the remainder, i.e.,
\begin{equation}
\lambda_{ij}=\lambda^{(s)}_{ij}+\lambda^{(b)}\delta_{ij}.
\end{equation}
Since the modifications must be made at fixed energy and particle densities, one also has the constraints
\begin{eqnarray}
\label{eq:erhoconstraints}
\Delta\rho_\ell&=&0=e^{-\alpha_\ell}\int\frac{d^3p'}{(2\pi)^3}e^{-E({\bf p}')/T}\left\{
-\Delta\alpha_\ell+\frac{E({\bf p}')\Delta T}{T^2}
\right\},\\
\Delta\epsilon&=&0=\sum_\ell e^{-\alpha_\ell}\int\frac{d^3p'}{(2\pi)^3}e^{-E({\bf p}')/T}\left\{
-\Delta\alpha_\ell E({\bf p}')+\frac{E({\bf p}')^2\Delta T}{T^2}-\frac{\lambda^{(b)}p'^2}{E({\bf p}')}
\right\}.
\end{eqnarray}
Using Eq. (\ref{eq:Adef}), one can substitute for $\lambda^{(s)}$ and $\lambda^{(b)}$ in the expressions for $\pi^{(s)}$ and $\pi^{(b)}$ in Eq. (\ref{eq:Tijprime}) and after using the constraints in Eq. (\ref{eq:erhoconstraints}) solve for $A^{(s)}$ and $A^{(b)}$. After some lengthy algebra,
\begin{eqnarray}
\label{eq:Aresults}
A^{(s)}&=&\frac{\Delta\tau}{2\eta},\\
\nonumber
A^{(b)}&=&\frac{\Delta\tau}{3\zeta},\\
\nonumber
\Delta T&=&\frac{\lambda^{(b)}PT^2}{D},\\
\nonumber
\Delta \alpha_\ell&=&-\frac{\lambda^{(b)}P\epsilon_\ell}{D\rho_\ell},
\end{eqnarray}
where $D$ is given in Eq. (\ref{eq:Ddef}), the ratios $\eta/\tau$ and $\zeta/\tau$ are given by Eq.s (\ref{eq:shear}) and Eq. (\ref{eq:zetaresult}). The results of Eq.s (\ref{eq:Aresults}) are consistent with expectations from the Navier Stokes equation. If the expressions for $\lambda$ are replaced with the product of the relaxation time and velocity gradient from Eq. (\ref{eq:lambdaguess}), then Eq. (\ref{eq:Adef}) can be equated to the Navier-Stokes equation once Eq.s (\ref{eq:Aresults}) are applied.

The method to consistently generate a phase space density for a thermal distribution at fixed energy densities, particle densities and at a given departure from equilibrium of the stress-energy tensor, $\pi_{ij}$, can be summarized as follows. First, one finds the temperature and chemical potentials corresponding to equilibrated energy density and particle densities. If $\pi^{(b)}$ is non-zero, one must then alter the temperature and chemical potential according to Eq. (\ref{eq:Aresults}). After generating a thermal distribution, each momentum is scaled by the matrix $\delta_{ij}+\lambda_{ij}$, where $\lambda_{ij}$ is given by Eq.s (\ref{eq:Aresults}) and (\ref{eq:Adef}).

\begin{figure}
\centerline{\includegraphics[width=0.5\textwidth]{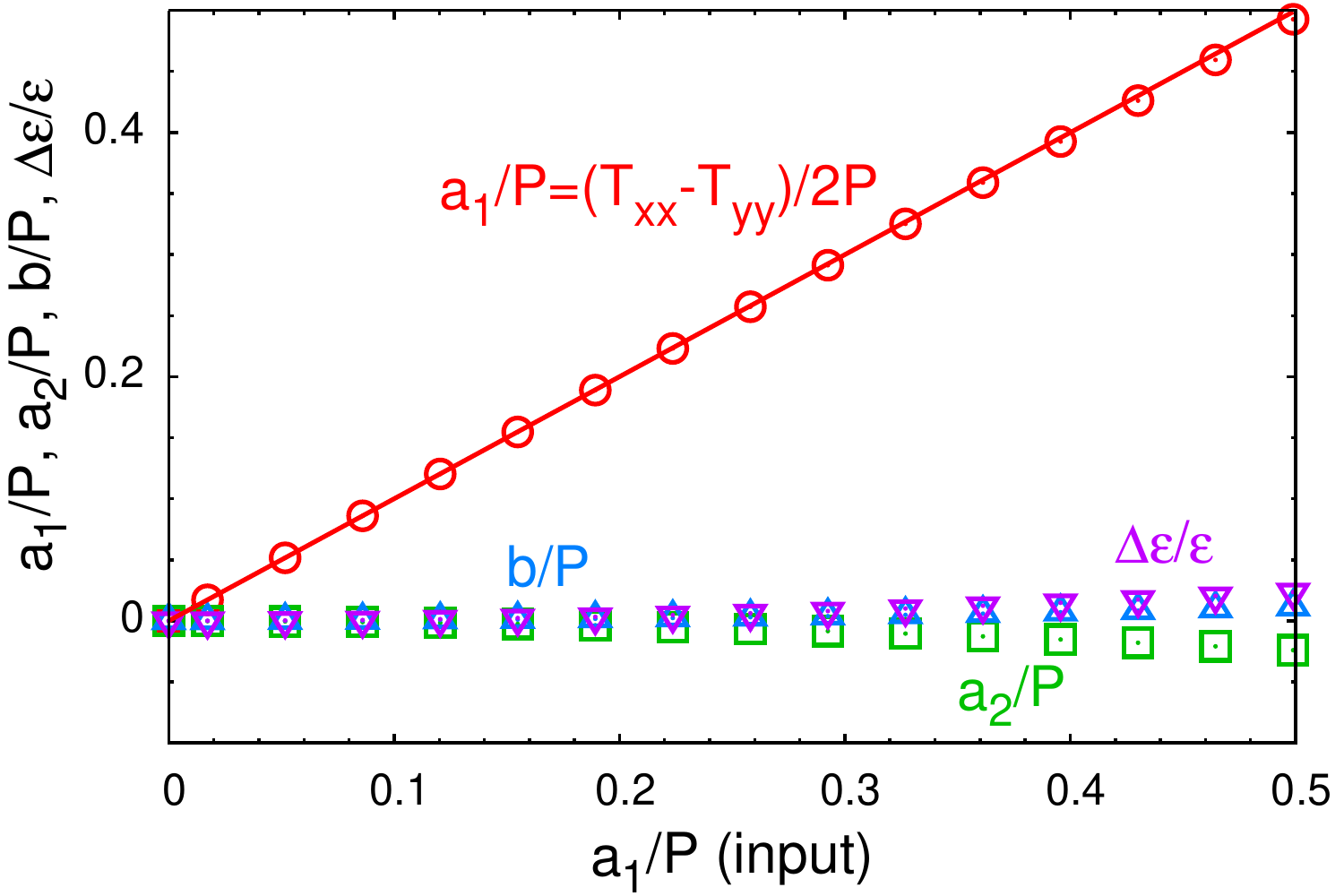}}
\caption{\label{fig:ltest}(color online)
For the procedure to reproduce a phase space density consistent with the stress energy density and particle densities, momenta were generated according to a thermalized distribution, then scaled linearly according to a matrix $\lambda_{ij}\propto\pi_{ij}$, and with the coefficient of proportionality chosen so that the procedure is exact for small input values, $\pi_{ij}^{\rm(input)}$. Here, the resulting values of the stress-energy tensor, as defined in Eq. (\ref{eq:abdef}), are shown as a function of the input values $a_1^{\rm(input)}$, which could come from a hydrodynamic simulation. For the test, the other input values are set to zero, i.e., $b^{\rm(input)}=a_2^{\rm(input)}=0$. The resulting value of $a_1$ (circles) closely follows the input value (line) over the same range. The other elements, $b$ (upward triangles), $a_2$ (green squares) and $\Delta\epsilon$ (downward triangles) stay near zero as desired for small $a_1^{\rm(input)}$, with relative deviations not staying below one percent for $a_1^{\rm(input)}/P\lesssim 1/3$. 
}
\end{figure}

The work above demonstrates that generating an equilibrated phase-space density, followed by a rescaling of the momenta using matrices built proportional to $\pi^{(s)}$ or $\pi^{(b)}$, one will generate stress-energy tensors consistent with the $\pi^{(s)}$ and $\pi^{(b)}$ used to scale the momenta. Further, the constants of proportionality relating $\lambda$ and $\pi$ can be found by integrating moments of the the thermal distributions, given in Eq.s (\ref{eq:Adef}). However, these relations were derived by keeping only the linear terms in $\lambda$ when calculating the stress-energy tensor in Eq. (\ref{eq:Tijprime}). The consistency from the above prescription should break down for large enough $\lambda$, or equivalently large enough $\pi^{(s)}$ or $\pi^{(b)}$, as the relation between $\lambda$ and $\pi$ will no longer be purely linear. To see the range to which the linear behavior extends, we consider an equilibrated gas of hadrons at a temperature of 160 MeV, with the phase space density modified according to the following form for $\lambda$,
\begin{equation}
\lambda_{xx}=-\lambda_{yy}=\lambda_{\rm mag}=A^{(s)}a_1^{\rm(input)},
\end{equation}
with all other $\lambda_{ij}=0$, and $a_1$ references one of the deviations to the stress energy tensor,
\begin{eqnarray}
\label{eq:abdef}
a_1&\equiv&(T_{xx}-T_{yy})/2,~~
a_2\equiv\frac{1}{\sqrt{12}}(2T_{zz}-T_{xx}-T_{yy}),\\
\nonumber
a_3&\equiv&T_{xy},~a_4\equiv T_{yz},~a_5\equiv T_{xz},\\
\nonumber
b&\equiv&\frac{1}{3}(T_{xx}+T_{yy}+T_{zz}-P)/3.
\end{eqnarray}
For the procedure outlined above to be successful the stress energy tensor generated with the specified $\lambda_{ij}$ should result in $a_1=a_1^{\rm(input)}$, where $a_1^{\rm(input)}$ refers to input value from the hydrodynamic simulation, whereas $a_1$ refers to result from integrating the phase space distribution. For the test, it is assumed that $a_{i\ne 1}^{\rm(input)}=0$ and $b^{\rm(input)}=0$, so that $a_{i\ne 1}$, $b$ and $\Delta\epsilon$ should all be zero. For small $a_1^{\rm(input)}$ the procedure should be exact, but for larger deviations non-linear contributions will lead to the output values varying from the input values. This is illustrated in Fig. \ref{fig:ltest}, which shows $a_i$, $b$ and the change in the energy density $\Delta\epsilon$ as a function $a_i^{\rm(input)}$. Given that systematic uncertainties, both theoretical and experimental, at RHIC are rarely below the 5\% level, any deviation of the order of one percent or less should be tolerable. As can be seen in Fig. \ref{fig:ltest}, the procedure described here appears sufficient for values of $a_1^{\rm(input)}/P\lesssim 1/3$. Although not shown here, similarly acceptable violations of the particle densities occur for large $a_1^{\rm(input)}/P\lesssim 1/3$. For larger deviations, one would have to either accept the discontinuities of the densities and of the stress-energy tensor, including the energy density, or apply a more sophisticated, and likely more difficult to implement, prescription.

Along the same lines as the tests illustrated in Fig. \ref{fig:ltest}, the validity of the linear considerations for correcting for bulk effects was also analyzed. Due to the small bulk viscosity for a gas, the input values of $b^{\rm(input)}/P$ are small, but nonetheless can require fairly large modifications of the phase space density since the linear coefficient $A^{(b)}\approx 1/\zeta$. Since $\lambda^{(b)}$ are not small, non-linear contributions tend to have similar corrections to the stress energy tensor as was the case for the shear viscosity. However, since the linear corrections are so small, the non-linear corrections quickly overwhelm the linear corrections and invalidate the prescription. Thus, while the linearity assumed in the procedure worked well for generating a consistent stress-energy tensor for the shear case even for fairly large values of $a_1^{\rm(input)}/P$, the linear procedure failed for the equivalent bulk case. For instance, if one were to set the viscosities according to the Navier-Stokes equation in a one-dimensional expansion, the magnitude of the components of $\lambda$ would be of the order of the relaxation time multiplied by the velocity gradient, for both the shear and bulk corrections. The linear procedure for correcting for both the shear and bulk terms would both lead to unwanted non-linear deviations of the energy and particle densities, and to the stress energy tensor. These deviations would be of the same order for both the shear and bulk corrections. However, for typical values of $\lambda$ seen in heavy ion collisions, the unwanted deviations to the stress-energy tensor tend to be much smaller than the input value $a_i^{\rm(input)}$, while they tend to be much larger than $b^{\rm(input)}$. Thus, there is little point to correcting for the bulk effects since the induced errors tend to be larger than the error associated with simply setting $b^{\rm(input)}=0$.

To generate particles according to the phase-space density and the evolution of the breakup hyper surface, one can apply the relation \cite{cf},
\begin{equation}
\label{eq:dN}
dN=\frac{f({\bf p})d^3p}{(2\pi)^3E({\bf p})}(p\cdot \Omega)\Theta(p\cdot \Omega),~~~
\Omega^\alpha\equiv \epsilon^{\alpha\beta\gamma\delta}\Delta X_\beta \Delta Y_\gamma \Delta Z_\delta. 
\end{equation}
The vectors $\Delta X$, $\Delta Y$ and $\Delta Z$ describe the widths of the ``volume'' element. For a surface element of the breakup hypersurface, defined by a quadraleteral, one can consider the corresponding surface element along the hypersurface at a later time. The three vectors describe the separation between the centers of each of the opposite faces. To understand how this provides a volume element, one can consider the hyper surface element defined simultaneously across the element, i.e., $\Delta X^0=\Delta Y^0=0$. The product $\epsilon^{\alpha\beta\gamma\delta}\Delta X_\gamma\Delta Y_\delta$ then becomes an antisymmetric matrix with the ``$i0$'' elements defining a vector perpendicular to the surface element whose magnitude is the area $\Delta A$. If $\Delta Z^0=0$, the boundary of the next surface element for the breakup surface is defined simultaneously with the previous one, and the product $\epsilon^{\alpha\beta\gamma\delta}\Delta X_\beta\Delta_Y\gamma\Delta Z_\delta$ is purely time-like with a magnitude equal to a volume $\Delta V$ bounded by the concurrent hyper-surface elements. The final convolution in Eq. (\ref{eq:dN}) thus gives $dN=f \Delta Vd^3p/(2\pi)^3$, as desired. If the hyper-surface element is defined at the same position, but different time, i.e., $\Delta Z=(\Delta\tau,0,0,0)$, the procedure yields $dN=f{\Delta{\bf A}\cdot p} d^3p/\epsilon_p$, which is what one expects for emission from a static surface. The step function $\Theta(p\cdot\Omega)$ in Eq. (\ref{eq:dN}) comes into play only for space-like  $\Omega^\alpha$, and eliminates those particles traveling inward into the surface \cite{csernai,bugaev}\footnote{This correction, of course, introduces some violation of conservation laws.  For realistic freeze-out conditions this violation is $\sim 5\%$, and going beyond it is somewhat non-trivial \cite{csernai}}.  For collisions at RHIC this amounts to only a few percent of all the particles. The step function must still be treated carefully, as the sign can depend on how $\Delta X$ and $\Delta Y$ are defined. For space-like $\Delta \Omega$, one should ensure that one is choosing those particles leaving, rather than entering, the breakup surface.

The algorithm for creating a sampling of particles from the element $\Delta\Omega$ using the description above is:
\begin{enumerate}
\item Calculate the ``volume'' vector $\Delta\Omega^\alpha$ as described above.
\item Assuming a real volume of size $V=|\Delta\Omega|$, calculate the number of particles $N$ one would create in a thermal system with that volume, temperature and chemical potential.
\item For each particle in $N$, create a momentum consistent with a static thermal distribution.
\item Scale the momentum according to $\lambda_{ij}$ determined from from the deviations of the stress-energy tensor as described in Eq. (\ref{eq:Adef}).
\item Boost the particle according to the velocity of the fluid element.
\item If $p\cdot\Omega<0$, throw out the particle. One must be careful to ensure that $p\cdot\Omega>0$ refers to particles leaving the element.
\item Keep or reject the particle according to the probability $p\cdot\Omega/E_p V$.
\end{enumerate}
It would be straight-forward to alter the procedure above to incorporate non-uniform relaxation times, i.e. $\lambda$ could be a function of the magnitude of ${\bf p}$ as measured in the fluid frame or could be a function of the species. The principal motivations for altering the form would be to incorporate longer relaxation times for high energy particles, given that they may require multiple collisions to re-thermalize, or to give longer relaxation times to species with longer mean free paths. It has been shown that the difference between saturating $v_2$ values for different species, often associated with quark-number scaling in coalescence, might partially derive from different relaxation times \cite{Dusling:2009df}. If two species are emitted from the same fireball, but characterized by different relaxation times, the species with the shorter relaxation time (e.g. protons) will have a more isotropic local momentum distribution. Since the anisotropy of the local phase space density lowers $v_2$, those particles with smaller relaxation times will then have higher values of $v_2$.

Comparing this procedure to the Grad form described by Eq. (\ref{eq:gradform}), reveals no difference if the departure from equilibrium is small, assuming an appropriate choice is made for $C(p)$ in Eq. (\ref{eq:gradform}). To see this, one can expand the argument in the exponential for the phase space density for small $\lambda$,
\begin{eqnarray}
f({\bf p}')&=&e^{-E({\bf p}')/T}\approx e^{-E({\bf p})/T}\left\{
1+\frac{p_i}{E_pT}\lambda_{ij}p_j\right\}\\
\nonumber
&=&e^{-E({\bf p})/T}\left\{1+A^{(s)}\frac{p_i}{E_pT}\pi^{(s)}_{ij}p_j\right\},
\end{eqnarray}
where the definition for $A^{(s)}$ in Eq. (\ref{eq:Adef}) was used to replace $\lambda$. By comparison with Eq. (\ref{eq:gradform}) one can see that the two expressions are identical for small $\pi$ if
\begin{equation}
C(p)=\frac{A^{(s)}}{E_pT}.
\end{equation}
For the assumption of uniform relaxation time applied thus far for the scaling procedure, $A^{(s)}$ was indpedendent of the momentum. However, one could in principle choose an arbitrary function form, i.e., instead of the assumption that the scaling function $\lambda$ is independent of momentum as in Eq. (\ref{eq:Adef}), one could instead have chosen,
\begin{equation}
\label{eq:lambdaFdef}
\lambda_{ij}(p)=F(p)A_F^{(s)}\pi_{ij}, 
\end{equation}
where $F(p)$ could be any arbitrary function of the magnitude of the momentum
(recent developments show $F(p)$ could indeed be non-trivial and species-dependent \cite{teamoore})
One would then retrace the procedure above and derive the value for $A_F^{(s)}$. In the small $\pi^{(s)}$ limit, the Grad form and the scaling procedure would then equate once one had chosen
\begin{equation}
\label{eq:Fdef}
C(p)=\frac{A^{(s)}_FF(p)}{E_pT}.
\end{equation}
Even if one were to use the Grad form rather than the scaling procedure presented here, this expression offers insight as to how one might choose the form for $C(p)$. If one thought the relaxation time were independent of momentum, or particle species, one would choose $C(p)$ consistent with $F(p)$ being constant. Otherwise, it would make most sense to choose $F(p)$ proportional to the relaxation time. For higher momentum hadrons, one might expect longer relaxation times, not so much because of reductions in the overall cross sections, but because it can take several collisions for a high energy particle to completely lose memory of its momentum. The extreme case would be a high-energy jet, for which the relaxation time might be so large that it cannot re-thermalize on the time scale of the expansion. In this limit both the Grad form and the scaling procedure lose their validity.

The Grad form, Eq. (\ref{eq:gradform}), has drawbacks in that it yields negative phase space densities for large momentum. The momentum range for which the phase space densities become unphysical depends on the size of the viscous corrections. For correction of the order $a_1/P\sim 1/3$, negative phase space densities ensue for momenta only a few times the thermal momentum. This problem occurs even for the case where $F(p)$ is chosen to be a constant. If one chooses a functional form $F(p)$ in Eq. (\ref{eq:Fdef}) that grows with momentum, the problems of negative phase space densities become all the more acute.

The scaling algorithm outlined above also has some drawbacks in that one loses the equivalence between the non-equilibrium components of the generated stress-energy tensor and the non-equilibrium components used to determine the coefficients in Eq. (\ref{eq:Adef}) is lost for large departures from equilibrium. One advantage of the procedure described above is that it accommodates Monte Carlo generation of particles in a fairly straight-forward manner. The procedure is also better behaved at large momentum. From Eq. (\ref{eq:lambdaFdef}), one can see that if all particles are given a fixed lifefime, i.e. $F(p)$ is a constant, that the scaling matrix $\lambda$ is independent of momentum, and as long as $\lambda$ is not large, the linear approximation is reasonable for all momenta. However, if $F(p)$ increases with momentum, there will again be some momentum range for which the approximation becomes unwarranted and perhaps even unphysical. 

Both procedures have difficulty adjusting for bulk corrections to the stress-energy tensor, i.e. $b^{\rm(input)}\ne 0$, as one must also adjust the temperature and chemical potentials to maintain fixed energy and particle number densities. Once these densities are conserved, both forms require large deviations of the phase space density to produce small $b/P$. Although it would be straight-forward to do this consistently, linear approximations (assuming that the modification factors are proportional to $\pi^{(b)}$), are bound to run into difficulties such as negative phase space densities for specific momenta in the Grad form, or for non-linear corrections overwhelming the linear corrections for the scaling procedure. Given that the bulk corrections to the stress energy tensor tend to be less than one percent of the pressure in the region of hydrodynamic/microscopic interface, the best course of action may be to ignore the bulk correction altogether. This recommendation would change if the bulk correction were not small for the hadron case. Such might be the case if the hadronic mean fields \cite{paech}, e.g. the chiral condensate, are away from equilibrium. However, that would suggest adding a dynamic mean field to both the hydrodynamic and microscopic prescriptions \cite{paechdumitru}, rather than attempting to incorporate the viscous corrections into the momentum distributions, which as shown in Sec. \ref{sec:bulk}, can accommodate only small viscous corrections.

\section{Summary}
By calculating the bulk and shear from both a dynamic perspective and from Kubo relations, it is clear how one should alter the phase space distribution for a system where the relaxation time is independent of species or of momentum. The calculations of the bulk viscosity were especially useful, as they illustrate why the bulk viscosity is so small for most applications to heavy ion collisions. In particular, one can see that $\zeta\rightarrow 0$ for both ultra-relativistic and non-relativistic system. A mixture of the massive ($T<<m$) and massless ($m<<T$) components leads to the largest values of $\zeta/P$, but even in those cases bulk viscosity from kinetic non-equilibrium is negligible for designing an interface between hydrodynamic and Boltzmann models in heavy ion collisions.

The viscous calculations alluded to above inspired the algorithm described in Sec. \ref{sec:interface} for interfacing hydrodynamic and microscopic descriptions. The method was based on generating particles with a thermal distribution, then scaling their momenta linearly with a matrix composed of a unit matrix plus a piece proportional to the non-equilibrium stress-energy tensor taken from the hydrodynamic calculation. The constant of proportionality was then expressed in terms of integrals of the thermal distributions. The method was shown to be better than one percent accurate as long as the non-equilibrium shear components of the stress-energy tensor remain $\lesssim$ one third of the equilibrium pressure. The method did not appear useful for accurately generating non-equilibrium deviations of the bulk pressure, although such deviations are so small they can be safely neglected. The method has an advantage over previously considered techniques to generate corrected phase space densities, in that the phase space densities remain positive for all momenta. The proposed method also lends itself well to generating Monte Carlo ensembles of discrete momenta.

\section*{Acknowledgments}
Support was provided by the U.S. Department of Energy, Grant No. DE-FG02-03ER41259. 
G.T. acknowledges the financial support received from the Helmholtz International
Center for FAIR within the framework of the LOEWE program
(Landesoffensive zur Entwicklung Wissenschaftlich-\"Okonomischer
Exzellenz) launched by the State of Hesse.   

We wish to thank the organizers, participants and sponsors (HIC4FAIR and NIKHEF) of the ``Flow and dissipation in ultrarelativistic Heavy Ion Collisions'' workshop in Trento, where the discussions leading to the current work started.

\end{document}